# Open-Source Coil Matching Toolbox for Magnetic Stimulation and Other Electromagnetics (COMATOSE)

Max Koehler and Stefan Goetz

**ABSTRACT**

**The coil in transcranial magnetic stimulation (TMS) determines the spatial shape of the electromagnetic field in the head, which structures are concurrently activated, and how focal stimulation is. Most of the readily available coils have been designed intuitively instead of systematic mathematical-physical optimization as there were no methods available at the time. Previous research however demonstrated that these coils are far from optimum, e.g., for pulse energy or efficiency, and leave substantial room for lots of improvements. Techniques for rigorous mathematical optimization have been developed but are only available to very few groups worldwide. This paper presents an open-source toolbox, COMATOSE, to change that situation and make these methods available to a wider community. It incorporates the fundamental formalisms and offers vector space decomposition as well as base mapping as an explicit forward method, which is computationally less demanding than iterative computational optimization but can also form the initial solution for a subsequent optimization run if desired.**

**INTRODUCTION**

As a central element of transcranial magnetic stimulation (TMS), the stimulation coil determines the spatial shape of the electromagnetic field in the head, which structures are concurrently activated, and how focal stimulation is. Most of the readily available coils are based on decade-old designs from times when the understanding of the relevant physical aspects was more rudimentary [1,2]. Practically all coils with wider commercial distribution have been designed through human intuition and trial and error instead of a systematic design of physical and physiological properties [3]. The available mathematical formalisms and software packages would not have allowed that. Previous research however demonstrated that these coils are far from optimum, e.g., for pulse energy or efficiency, and leave substantial room for lots of improvements [4–10].

Systematically optimization or mathematical generative design can calculate coils out of objectives and constraints [11,12] or forward projection in combination with vector space decomposition (VSD) [13]. However, these techniques have only been available to very few groups worldwide who sufficiently master the mathematics and physics to design their own internal software packages. The open-source toolbox COMATOSE wants to change that situation and make these methods available to a wider community. It incorporates the fundamental formalisms and offers vector space decomposition as well as base mapping as an explicit forward method, which is computationally less demanding than iterative computational optimization but can also form the initial solution for a subsequent optimization run if desired. Furthermore, it provides all elements for modal optimization.

**BACKGROUND**

Basis vector representations and basis changes for magnetic fields has its origins in theoretical electromagnetism and quantum field theory to generally describe the entire solution space and include already all constraints to that finding a specific solution only requires finding coefficients, essentially, which share of each basis vector is needed. If the basis vectors are sorted in a sequence, e.g., with increasingly finer structure size, the identification of a solution can concentrate on just a finite number of basis vectors and ignore the rest due to their unpractically small features. The method has been applied to other fields in medicine, such as magnetic resonance imaging fields to describe and derive gradient as well as shim coils with specific features [14]. It enables a compact description of both desired and undesired field contributions and provides a direct link between coil geometry, current distributions, and resulting field modes, which represent the basis vectors of a vector space.

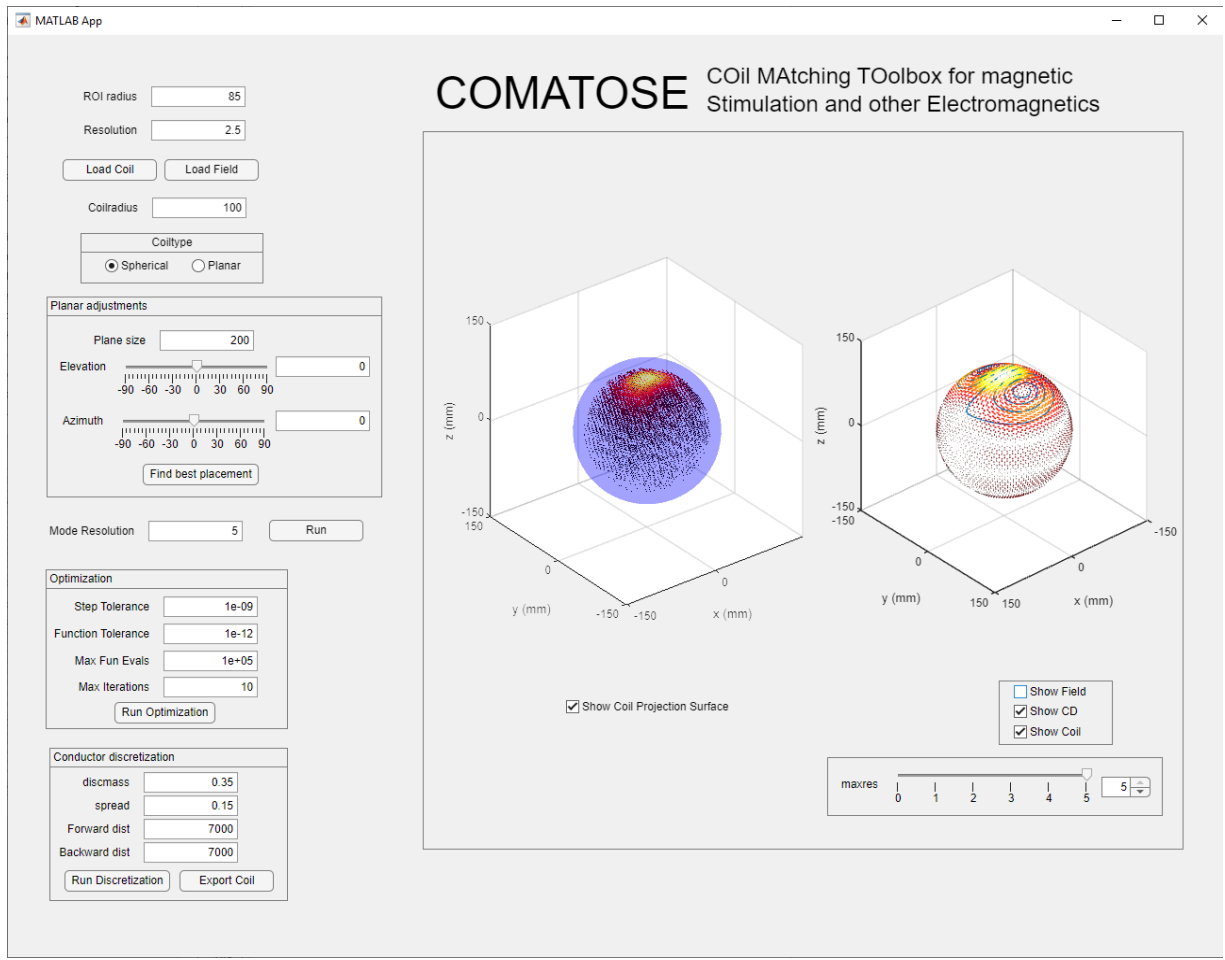

Figure 1. Main window of COMATOSE. On the left side, all environment parameters can be set or chosen, and procedures executed. The leftmost figure on the right shows the field-to-match and the coil projection surface (spherical in this case). The rightmost figure can show the things from the dropdown menu (resulting field, resulting current density and discretized wire path). With the slider below that figure, the resolution of the projection and therefore the projection outcome can be limited.

For TMS, superpositions of solenoidal modes, i.e., a weighted sum of basis vectors, can map out the solution space of all possible coils, e.g., on a flat or alternatively any other surface, e.g., some form of a bent, curved shape around the head, and already enforce all constraints [8]. Thus, all possible combinations represent a valid coil, and any possible coil can be represented. Furthermore, modes sorted with increasing spatial frequency, i.e., smaller and smaller features, allows concentrating on a finite number of modes. The modal representation of a coil through coefficients has similarities with the representation of a function as a sum of sinusoidal base vectors in a Fourier series but expand it essentially to more-dimensional space. Likewise, higher frequencies can often be neglected for practicality to concentrate on a finite number of elements. Finding the coefficients, i.e., the share of the various modes in the desired coil can use mapping, which mathematically represents a basis change and/or optimization.

Once, the shares of the various modes to form a solution are found, the coil character is widely described. The specific winding with discrete wire, however, is still missing. Discretization of the continuous vector field at the end of the mapping or optimization generates a concentrated wire path. Mathematically, this step preferably forms an integral curve through the coil current distribution field; an overlaid gradient field generates the spiral [13]. This procedure eliminates previous ad-hoc methods and heuristics such as connecting closed loops as a consequence of solenoidal modes [15]. The discretization determines the inductance of the coil through the size of the gradient field.

The toolbox includes all elements and code to remap a coil onto a different surface, may it be flat or curved, and reduce the pulse energy and coil heating for the same spatial field profile in the head. Furthermore, it can generate coils to field shapes and targets that have not been represented by any coil before.

**APPLICATION**

We developed the toolbox COMATOSE to solve limitations in existing coil design tools and break the situation that only a few groups have access to these techniques right now. We implemented it as a MATLAB-based open-source application to allow coil optimization and magnetic field matching with a user-friendly graphical user interface (GUI). The toolbox implements vector space decomposition, base vector mapping, and optimization. The former is computationally more efficient if an existing coil design or target should be redesigned into a different, e.g., more efficient, smaller, or otherwise more practical coil [13].

The workflow of the application comprises four distinct steps. In the first step, users define the simulation environment by setting key parameters such as the *region of interest (ROI) radius*, which approximates the brain as a spherical region and the *resolution*, which determines the density of sampling points within the ROI. Subsequently, users load the magnetic vector potential or coil geometry to be analyzed. For predefined magnetic vector potentials, the software provides options for adjusting the field's position along the x-, y-, or z-axes. Alternatively, an existing coil can be loaded, and its corresponding magnetic vector potential is calculated, a feature comparable to previously described tools such as the TMS coil design instrument [16].

The second step defines the coil projection surface. At present, COMATOSE allows users to choose between spherical and planar coil geometries. As the toolbox is open-source, users are invited to include other shapes. For planar coils, users can adjust azimuth and elevation manually, or the application can automatically align the projection surface to the target magnetic field using the *find best placement* feature. Users also specify the *mode resolution*, which determines the degree and order of spherical harmonics for spherical coils. For planar coils, it governs the resolution, i.e., the number of vortices, along the x- and y-axes of the coil projection surface.

In the third step, the application computes the current densities on the chosen coil projection surface, the corresponding magnetic vector potentials in the ROI, and VSD coefficients based on the specified projection surface. Users initiate these calculations by clicking the *run* button. The toolbox visualizes the results graphically for further evaluation. Through an interactive slider, users can alter the maximum resolution to refine the projected magnetic field. For users seeking additional capabilities, an optional optimization step can be performed on top of the projection results. Optimization parameters can be configured through the GUI to meet specific requirements.

The fourth and final step discretizes the resulting current density into conductor paths to make it usable in downstream applications, e.g., computer-aided design (CAD) for manufacturing. Several parameters control the discretization process: *Discmass* introduces an optional inertia-like effect to discretization, which can suppress extremely small structures and bends in the conductor path. *Spread* determines the factor of the overlaying gradient field and therefore the number of turns as well as the inductance of the resulting coil. *Forward* and *backward distance* set the length of the coil in either direction, as the discretization starts in the point of highest current density.

Figure 1 visualizes on the left side the targeted spatial profile of the magnetic vector potential inside a spherical ROI. The right plot represents the calculated matched coil current density with a possible discretized conductor path on top. The coil winding shape in the form of the conductor path can be exported as comma-separated values (CSV).

## Conclusion

COMATOSE offers several functions which we observed to be available to few groups only and risk to hamper further technological progress. It enables users to redesign and simplify existing coils with improved efficiency, create novel coil geometries tailored to specific applications, and seamlessly export conductor paths for direct use in CAD software. The tool also includes forward mapping based on vector space projection for low computational demand and iterative computational optimization.

### Acknowledgement

The authors are inventors on patents and patent applications on brain stimulation technology, independent from this work.

## Availability

The application is available in the MATLAB Add-Ons Manager under the name COMATOSE and on GitHub (https://github.com/maxxkoehler/COMATOSE). The GitHub repository offers the toolbox as an application to be opened in MATLAB or as a precompiled standalone application, that just need a MATLAB runtime, which can be downloaded free of charge.